\documentclass[aps,prl,twocolumn,floatfix]{revtex4}
\usepackage{graphicx}
\usepackage{dcolumn}
\usepackage{bm}
\usepackage{amsmath}
\usepackage{epstopdf}
\usepackage{hyperref}

\usepackage{epstopdf, epsfig}
\usepackage{textcomp}
\usepackage{color}

\newcommand{\be}{\begin{equation}}
\newcommand{\ee}{\end{equation}}

\usepackage{blindtext}
\usepackage{lettrine}

\begin{document}


\title{Pattern selection and restricted vortex dynamics by spatial periodic forcing in rapidly rotating Rayleigh-B\'enard convection}

\author{Shan-Shan Ding}
\author{Hong-Lin Zhang}
\author{Dong-Tian Chen}
\author{Hao-Han Sun}
\author{Jin-Qiang Zhong}
\email{jinqiang@tongji.edu.cn}

\affiliation{School of Physics Science and Engineering, Tongji University, Shanghai 200092, China}

\date{\today}

\begin{abstract}
Pattern forming with externally imposed symmetry is ubiquitous in nature but lightly studied. We present experimental studies of pattern formation and selection by spatial periodic forcing in rapidly rotating convection.    
We observe symmetric convection patterns in form of regular vortex lattice near the instability onset, when the periodicity of the external forcing is set close to the intrinsic vortex spacing.
We show that the new patterns arise as a dynamical process of imperfect bifurcation which can be well described by a Ginzburg-Landau-like model.
With increasing buoyancy strength the effect of external forcing weakens, and the convective vortices evolve from a stationary state to exhibit restricted and finally stochastic motions. 
\end{abstract}

\maketitle

Pattern formatting phenomena are omnipresent in a wide variety of physical, chemical, and biological systems \cite{CH93,GL99,ZZ70,FB85,JD89}. Nonequilibrium spatiotemporal patterns often arise through symmetry-breaking bifurcations when an initially uniform system is driven internally away from thermodynamic equilibrium. Many natural systems are, however, often constrained with nonuniform boundaries having broken symmetry that may modify the pattern formatting process and influence the intrinsic flow organization. Examples include atmospheric convection rolls forming over mesoscale topography \cite{TP03}, formation of Taylor columns over seamounts that determine the overlying pattern of ice-cover in high-latitude oceans \cite{MD97}. Exploring the interaction between externally imposed symmetries and intrinsic symmetries preferred by the system may shed new light onto the complexity in pattern formatting process, and enable us to induce, control or eliminate patterns using external forcing in various systems \cite{KP78,LGL83,Co86,DBZE01,IRGSW01,Mc07,SWMPB08}.

The fundamental physics of pattern formation have been studied over the past few decades in carefully controlled experimental systems, such as in rotating Rayleigh-B\'enard convection (RBC) \cite{CH93,ZES91,BLNA98}, i. e., a fluid layer heated from below and rotated about a vertical axis with angular velocity $\Omega_\mathrm{D}$. When the temperature difference ${\Delta}T$ exceeds the onset ${\Delta}T_\mathrm{c}$($\Omega_\mathrm{D}$), convection pattern appears first as straight rolls under slow rotations \cite{KL69,HEA97,BAP02}, which becomes unstable to the Kuppers-Lortz instability when $\Omega_\mathrm{D}$ increases \cite{HPAE98,SSC05}. 
Square (or hexagonal) patterns may form when the dimensionless rotation rate, $\Omega{=}\Omega_\mathrm{D}H^2/\nu$ ($\nu$ is the kinematic viscosity), reaches the order of 100 \cite{BLNA98}. With further increase in $\Omega$, 
the flow pattern becomes disordered \cite{BAP02}.  It is observed that under large rotation rates ($\Omega{\ge}10^4$) the flow field near onset is characterized by columnar vortices that exhibit Brownian-type random motion \cite{NTYM19, Paper1, Paper2}. 
Clearly in rapidly rotating convection the spatiotemporal periodicity of the flow pattern is lost even close to onset. 
Previous studies have shown that when flow patterns are modulated by a spatially periodic perturbation, a commensurate state often arises in which the structural periodicity of the flow rationally accommodates to that of the perturbation, as a result of competition between the imposed and intrinsic patterns \cite{LGL83,IRGSW01,SWMPB08}. 
It raises naturally the intriguing question whether flow patterns with spatial periodicity and symmetry can be restored in rapidly rotating convection when external forcing is applied.

In this Letter we report experimental observation of orderly flow patterns forming under the control of externally imposed topographic forcing in rapidly rotating convection. We use a convection apparatus that was designed for high-resolution flow structure measurements in rotating RBC \cite{Paper1, Paper2, SLDZ20}. We use a cylindrical cell that had a diameter $d{=}235.0$ mm and a height $H{=}63.0$ mm, yielding an aspect ratio $\Gamma{=}d/H{=}3.7$. The bottom plate of the sample, made of oxygen-free copper, is finely machined to construct an array of thin cylinders extending out from the plate surface. These raised cylinders are periodically spaced to form a square- or hexagon-patterned surface. 
Flow patterns at a fluid depth of $z{=}H/4$ are measured using the technique of particle image velocimetry (see Supplemental Material \cite{SM} for detailed experimental methods). The experiment is conducted with a constant Prandtl number Pr${=}\nu/{\kappa}{=}4.38$ and in the range $2.0{\times}10^6{\le}$Ra${\le}1.0{\times}10^8$ of the Rayleigh number $\mathrm{Ra}{=}{\alpha}g{\Delta}TH^3/{\kappa}{\nu}$. Here $g$ is the gravitational acceleration, $\alpha$, $\kappa$ are respectively the isobaric thermal expansion coefficient, thermal diffusivity of the fluid. 
Rotating angular velocity of $0.6{\le}{\Omega_\mathrm{D}}{\le}2.0$ rad/s are used thus $3.6{\times}10^3{\le}{\Omega}{\le}1.2{\times}10^4$. The reduced Rayleigh number, $\varepsilon{=}(\mathrm{Ra}{-}\mathrm{Ra}_\mathrm{c})/\mathrm{Ra}_\mathrm{c}$, spans the range ${-}0.6{\le}{\varepsilon}{\le}13.5$. 
Here $\mathrm{Ra_c}{=}\mathrm{Ra}({\Delta}T_\mathrm{c})$ is the onset of convection \cite{Onset}. 
The Froude number, $\mathrm{Fr}{=}{\Omega}_\mathrm{D}^2d/2g$, covers the range $4.4{\times}10^{{-}3}{\le}\mathrm{Fr}{\le}0.05$.  

\begin{figure}
\includegraphics[width=0.5\textwidth]{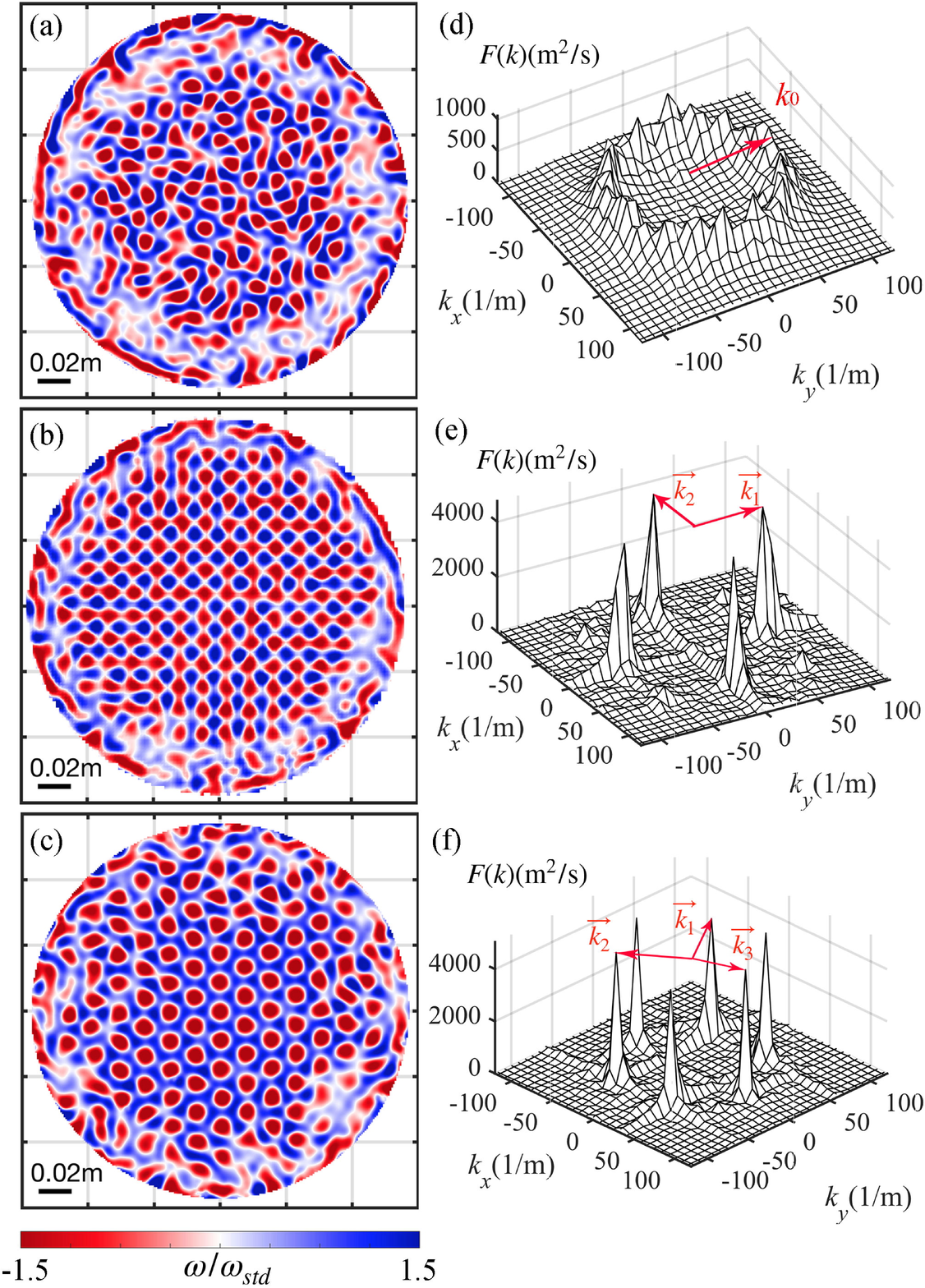}
\caption{(a-c) Instantaneous vertical vorticity distribution $\omega/\omega_{std}$. $\omega_{std}$ is the standard deviation of $\omega$. Results are for the reference cell (a), the square-patterned (b) and the hexagon-patterned (c) cells. $\Omega{=}1.12{\times}10^4$ and $\varepsilon{=}0.49$. The distances between adjacent cylinders are $\lambda{=}14.14$ mm (b) and $17.32$ mm (c). (d-f) Fourier spectra $F(\vec{k})$ of the vorticity field, determined by $\omega(\vec{r})$ in the central region of $60{\times}60$ mm$^2$ shown in (a-c) respectively. The arrow in (d) shows the mean radius $k_0$ of the crater-like structure. The arrows in (e) and (f) represent the characteristic wave vectors $\vec{k}_\mathrm{f}{=}\vec{k}_i (i{=}1,2...)$ of the imposed textures. Movies for (a-c) are available \cite{SM}.}
\label{fig:1} 
\end{figure} 

Figure 1 presents the flow patterns at the measured fluid height with $\Omega{=}1.12{\times}10^4$ and $\varepsilon{=}0.49$. When a flat bottom plate is used (i. e., the reference cell without external forcing), the flow fields of the vertical vorticity $\omega(\vec{r})$ are characterized by columnar vortices, which exhibit stochastic horizontal motions as reported in previous studies \cite{NTYM19, Paper1, Paper2}. Despite their random motion, the vortices maintain approximately a constant distance $\lambda_0{\approx}13.15$ mm with their neighboring vortices (Fig.\ 1a).
In the spatial Fourier spectrum $F(\vec{k})$ of vorticity field calculated in central region (Fig.\ 1d), a crater-like structure with a radius $k_0{=}1/\lambda_0$
is apparent in the spectrum $F(\vec{k})$, indicating that the vortices are distributed with random orientations but with a preferred spacing. When periodic topographic structures are constructed on the bottom plate, they modulate both the local temperature and the shearing interaction of the fluid with the solid surface, leading to new convection patterns. Figures 1b and 1c show the vorticity fields when the bottom surface is textured with a square and a hexagonal array of cylinders, respectively, with their spacing $\lambda$ chosen close to the intrinsic wavelength of the vorticity field $\lambda_0$. 
Since the fluid overlying the raised cylinders is relatively hotter, upwelling vortices (i.e., cyclones when observed in the lower half fluid layer \cite{Note}) tend to form above the cylinders, forming a 1:1 commensurate structure with respect to the bottom texture. The downwelling vortices (anticyclones), however, appear in between the raised cylinders. In the square-patterned cell (Fig.\ 1b), both the cyclones and anticyclones constitute a regular square lattice. 
The flow pattern induced by a hexagonal array of cylinders (Fig.\ 1c), however, consists of a hexagonal lattice of anticyclones with a cyclone located at the hexagon center. 
Such patterns are stationary and persist during the experiment.
Figures 1e and 1f present the Fourier spectra of the vorticity fields in the central region of Figs.\ 1b and 1c, respectively. In these spectra we see clear peaks located precisely at the wave vectors  $\vec{k}_\mathrm{f}{=}\vec{k}_i (i{=}1,2...)$ of the periodically imposed textures. These peaks of $F(\vec{k})$ are all very sharp and their amplitudes are approximately equal, suggesting that regular patterns with prescribed periodicity and symmetry are developed. 
Near the sidewall region ($r{\ge}100$ mm) where the imposed texture is absent, the flow field is time-varying and the vortex dynamics therein is largely influenced by the retrogradely traveling boundary zonal flow \cite{WGMCCK20, ZvHWZAEWBS20}.

\begin{figure}
\includegraphics[width=0.5\textwidth]{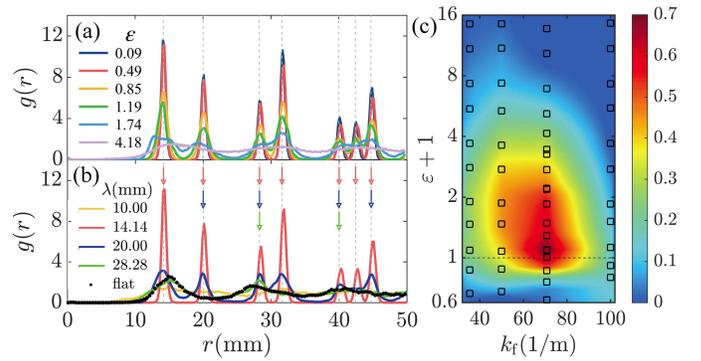}
\caption{(a, b) Radial distribution function $g(r)$ of cyclones in the cell textured with a square array of cylinders. Results are for various $\varepsilon$ with a constant cylinder spacing $\lambda{=}14.14$ mm in (a), and for various $\lambda$ with $\varepsilon{=}0.49$ in (b). The red, blue and green arrows in (b) denote the main and subharmonic wavelengths $r_{ij}$ of the imposed pattern for $\lambda{=}14.14, 20.00$ and $28.28$ mm, respectively. Dotted line in (b): result of the reference cell. (c) Contour plot of cross-correlation coefficient $C$ of the vorticity field and bottom texture in the $k_{f}{-}\varepsilon$ phase diagram. Open symbols represent the data points measured in the square-patterned cell; color contours are estimated from interpolation between these points. The horizontal dashed lines denote the onset $\varepsilon{=}0$. Results are for $\Omega{=}1.12{\times}10^4$.}
\label{fig:2} 
\end{figure} 

The observed spatial pattern can be quantified by the radial distribution function $g(r)$ of the vortices, which is defined as the ratio of the actual number of cyclones lying within an annulus region of $r$ and $r{+}{\delta}r$, to the expected number for uniform vortex distribution \cite{Paper1}. Figure 2a shows $g(r)$ for cyclonic distribution for various $\varepsilon$ in the square-patterned cell with $\lambda{=}14.14$ mm. Near onset ($\varepsilon{=}0.09$) multiple sharp peaks appear in $g(r)$, which locate at distances that match with the main and subharmonic wavelengths $r_{ij}$ of the forced square pattern at the bottom plate, fulfilling the condition: $r_{ij}{=}{\lambda}\sqrt{i^2{+}j^2}$, $(i,j{=}0,1,2...$ and $i{+}j{\ge}1)$. 
With increasing $\varepsilon$, the peak amplitudes in $g(r)$ decrease while the peak widths increase, signifying a less regular flow pattern. The multiple-peak structure is eventually flattened and $g(r)$ becomes close to a uniform distribution for $\varepsilon{\ge}4.0$, where we see the vortices exhibit apparent horizontal motions.
We examine as well the role of the periodicity of external forcing on the convection pattern. Figure 2b presents results of $g(r)$ near onset ($\varepsilon{=}0.49$) for various cylinder spacing $\lambda{=}10.00, 14.14, 20.00, 28.28$ mm. 
We see multiple peaks still appear at the main and subharmonic wavelengths $r_{ij}(\lambda)$. For $\lambda{\ge}20.00 $mm, the first peak is found near $r{=}\lambda_0$ which is associated with the intrinsic wavelength of the flow field. When $\lambda$ is far apart from $\lambda_0$, the maxima of $g(r)$ become less dominant and $g(r)$ approaches the result of the reference cell. 

The degree of matching between the flow pattern and the bottom texture can be evaluated through the cross-correlation coefficient $C$ of the vorticity field and the bottom texture, defined as 
$C{=}{\langle}\Delta{\omega}(\vec{r}){\cdot}\Delta{M}(\vec{r}){\rangle}/\sqrt{({\langle}\Delta{\omega}(\vec{r})^2\Delta{M}(\vec{r})^2{\rangle})}$, 
where $\Delta{\omega}(\vec{r}){=}{\omega}(\vec{r}){-}{\langle}{\omega}{\rangle}$, $\Delta{M}(\vec{r}){=}M(\vec{r}){-}{\langle}M{\rangle}$, and $M(\vec{r}){=}0$ (or ${-}1$) for the flat (or raised) area, representing the bottom surface profile.
${\langle}{\cdot}{\rangle}$ denotes a spatial average. 
Figure 2c summarizes the results of $C$ in the square-patterned cell for varying $\varepsilon$ and wave vector $k_\mathrm{f}{=}{\vert}\vec{k}_1{\vert}{=}{\vert}\vec{k}_2{\vert}$ of the external forcing. 
In this phase diagram, we see that $C(k_\mathrm{f},\varepsilon)$ has a single maximum ($C{\approx}0.7$) occurring at ($k_\mathrm{m}{=}70.7 \mathrm{m}^{-1}, \varepsilon_\mathrm{m}{=}0.09$), which implies the optimal conditions for pattern selection. 
In the vicinity of $(k_\mathrm{m}, \varepsilon_\mathrm{m})$, the spatial distribution of the vortices severely conforms to the bottom texture. The coefficient $C$ decreases if the control parameters $(k_\mathrm{f}, \varepsilon)$ deviate from $(k_\mathrm{m}, \varepsilon_\mathrm{m})$. The decreasing of $C$ with increasing $\varepsilon$ is slowest when a near-resonant external forcing $(k_\mathrm{f}{=}k_m{\approx}k_0)$ is chosen. The convection pattern and the imposed texture becomes essentially uncorrelated (with $C{\le}0.1$) when $(k_\mathrm{f}, \varepsilon)$ are set apart from $(k_\mathrm{m}, \varepsilon_\mathrm{m})$.  
Interestingly, when $k_\mathrm{f}{\approx}k_\mathrm{m}$, $C$ remains well above zero for $\varepsilon{<}0$, suggesting that under external forcing convection sets in with finite amplitude in the subcritical regime. 

We measure the time-averaged vorticity modulus ${\langle}{\vert}{\omega}{\vert}{\rangle}$ in an area of $65.8{\times}54.8$ mm$^2$ at the center of the cell, while slowly scanning ${\Delta}T$ in the range $-0.4{\le}\varepsilon{\le}0.4$. Results of ${\langle}{\vert}{\omega}{\vert}{\rangle}(\varepsilon)$ for the reference cell and two forced cells are shown in Fig.\ 3. 
Overall, these data suggest two distinct types of bifurcations when $\varepsilon$ increases and crosses zero. The reference cell data reveal a sharp transition from a non-convection state with ${\langle}{\vert}{\omega}{\vert}{\rangle}{=}0$ for $\varepsilon{\le}0$, to a convection state in which ${\langle}{\vert}{\omega}{\vert}{\rangle}$ increases rapidly for $\varepsilon{>}0$. 
For both forced cells, however, ${\langle}{\vert}{\omega}{\vert}{\rangle}$ remains positive for $\varepsilon{\ge}{-}0.4$ and grows relatively slowly with increasing $\varepsilon$, suggesting a smooth transition. The three inset panels present the vorticity fields captured in the three cells with approximately the same subcriticality ($\varepsilon{\approx}{-}0.1$). They demonstrate that while the fluid is still quiescent in the reference cell, apparent square (hexagonal) lattice of convective vortices have formed in the forced cells.   

\begin{figure}
\includegraphics[width=0.5\textwidth]{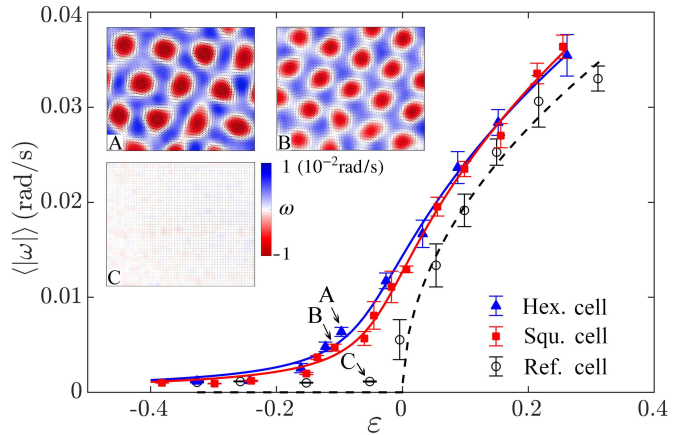}
\caption{Bifurcation curves of ${\langle}{\vert}{\omega}{\vert}{\rangle}(\varepsilon)$ near onset. The circles represent the data of the reference cell. The squares and triangles represent respectively data for the square-patterned cell with $\lambda{=}14.14$ mm, and the hexagon-patterned cell with $\lambda{=}17.32$ mm.  Error bars denote the standard deviation. The dotted line shows the fitted square-root law for the reference cell, and the solid lines are the predicted imperfect bifurcation curves for the forced cells. Inset panels: time-averaged vorticity fields for $\varepsilon{\approx}{-}0.1$ of the three cells. Data for $\Omega{=}1.12{\times}10^4$.}
\label{fig:3} 
\end{figure} 

In an effort to understand the $\varepsilon$-dependence of ${\langle}{\vert}{\omega}{\vert}{\rangle}$ near onset, we propose a phenomenological Ginzburg-Landau-like model for the convection amplitude $A_j$ of rotating RBC \cite{Sc07,SMK10} in the presence of external periodic forcing \cite{KP78,Co86,Mc07,SWMPB08}
\be
{\partial}_tA_j{=}{\varepsilon}A_j{+}{\xi}_0^2{\nabla}^2A_j{-}\sum_{i{=}1}^ng_0^{ij}|A_i|^2A_j{+}g_2^j{\delta}_jA_j^{{\ast}m{-}1}.
\ee
Here $g_0^{ij}$ is the nonlinear coupling coefficient between the Fourier modes $i$ and $j$ of the flow pattern and $n$ is the number of dominant modes.  $g_2^j$ is an imperfection coefficient. $\delta_j$ represents the strength of the external forcing. The integer $m$ denotes the degree of resonance and we consider here resonant forcing ($k_\mathrm{f}{\approx}k_0$ and $m{=}1$). $\xi_0$ is a spatial gradient coefficient.
In view of the symmetry of the flow patterns shown in Figs.\ 1b and 1c, we set $g_0^{ij}{=}g_0$ for any two coupling modes as constants and $g_2^j{=}g_2$. We consider a stationary solution for the near-onset flow regime. Through summation of all modes Eq. (1) is reduced to an amplitude equation: 
${(\varepsilon{+}\varepsilon_0)}A{-}g_0{\vert}A{\vert}^2A/n{+}g_2{\delta}{=}0$, 
with $\delta{=}\sum_{j{=}1}^n{\delta}_j$. The offset $\varepsilon_0$ represents a shift of $\mathrm{Ra}_\mathrm{c}$ since strictly $\vec{k}_\mathrm{f}{\ne}\vec{k}_0$ \cite{Cr80}, and since the local temperature gradient increases slightly over the bottom texture \cite{Mc07,SWMPB08}. 
The amplitude $A{=}\sum_{j{=}1}^nA_j$, is related to the observed mean vorticity modulus ${\langle}{\vert}{\omega}{\vert}{\rangle}$ through a scale factor $S$, $A{=}S{\langle}{\vert}{\omega}{\vert}{\rangle}$ \cite{SM}. 
We show in Fig.\ 3 the theoretical predictions ${\langle}{\vert}{\omega}{\vert}{\rangle}{=}\sqrt{\varepsilon/g_0}S^{{-}1}$ for the reference cell ($\delta{=}0$), and ${\langle}{\vert}{\omega}{\vert}{\rangle}(\varepsilon)$ for the two forced cells that fulfills a cubic equation. Both the experimental data and the theoretical curve show clearly the signature of a forward bifurcation near $\varepsilon{=}0$ when the external forcing is absent. Meanwhile the pronounced rounding of the transition in the two forced cells indicates an imperfect bifurcation. 
The parameter values obtained from fitting the bifurcation data can be interpreted as well by the model \cite{SM}. The agreement between the experimental and theoretical results suggest that the physics of external modulation near onset of rotating convection can be well described by the present Ginzburg-Landau-like model.


\begin{figure}
\includegraphics[width=0.5\textwidth]{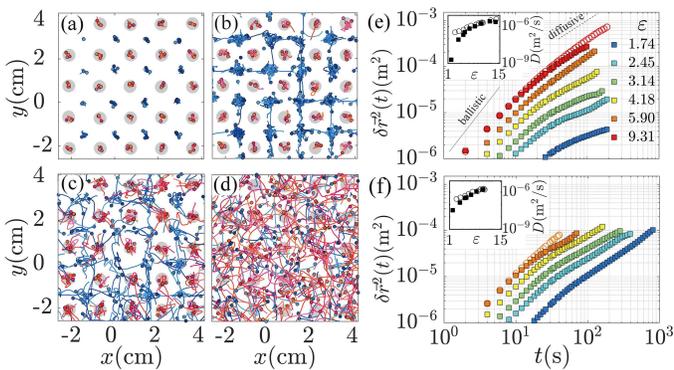}
\caption{(a-d) Trajectories of cyclones (red) and anticyclones (blue) overlying a square array of cylinders (gray circles). The dark (light) circle on each trajectory indicate where the vortex appears (terminates). Results for $\Omega{=}1.12{\times}10^4$, $\lambda{=}14.14$ mm and $\varepsilon{=}0.49$ for (a), 1.74 (b), 4.18 (c) and 9.31 (d). (e, f) Mean-square-displacements ${\delta}\vec{r}^2(t)$ of cyclones (e) and anticyclones (f) in the square-patterned cell. Open circles show the data of the reference cell. Insets in (e) and (f) show the diffusion coefficient $D$ of vortex motion in the reference cell (open circles) and the forced cell (solid squares).}
\label{fig:3} 
\end{figure} 

The convection pattern becomes time-dependent and the vortices are meandering horizontally when $\varepsilon$ is far above the onset. To explore the impact of the bottom texture on the flow field far from onset, we examine here the motions of the vortices with varying buoyancy strength. Figures 4a-4d show the trajectories of the vortices, overlying a gray texture that represents a square array of cylinders on the bottom surface. 
Near onset ($\varepsilon{=}0.49$), the motion of cyclones is severely restricted within the fluid domain over the raised cylinders. Meanwhile anticyclones appear at their stable positions, i.e., the middle points of every four adjacent cylinders, so as to maintain the largest distance from its nearest cyclones (Fig.\ 4a). 
With increased buoyancy forcing ($\varepsilon{=}1.74$), while the cyclones are still confined above the raised cylinders, the anticyclones may move in between the stable positions, with their chaotic trajectories forming a web-shape domain of anticyclonic motion (Fig.\ 4b). 
A further increase in $\varepsilon$ ($\varepsilon{=}4.18$) releases the constraint of vortex motion as shown in Fig.\ 4c. 
When a sufficiently large buoyancy forcing ($\varepsilon{=}9.31$) is applied, the trajectories of both cyclones and anticyclones cover nearly the full fluid domain as the effect of external forcing diminishes (Fig.\ 4d), and we find that the stochastic vortex motion \cite{Paper1,Paper2} recovers.    

To characterize the statistical behavior of the vortices, we compute their mean-square displacements, ${\delta}\vec{r}^2(t){=}{\langle}(\vec{r}(\tau{+}t){-}\vec{r}(\tau))^2{\rangle}_{\tau}$, using the trajectories $\vec{r}(t)$ obtained above. Figures 4e and 4f show the results for cyclones and anticyclones, respectively. 
With strong buoyancy forcing the data curves of ${\delta}\vec{r}^2(t)$ for cyclones in the forced cell resemble that of the reference cell (Fig.\ 4e), which reveal that the vortex motion undergoes a transition from ballistic to diffusive motion as time increases \cite{Paper1}. 
When $\varepsilon$ decreases ${\delta}\vec{r}^2(t)$ of the forced cell decreases progressively, signifying that the restriction effect of the cyclonic motion due to the external forcing becomes significant. A similar $\varepsilon$-dependence of ${\delta}\vec{r}^2(t)$ is found in the anticyclonic data (Fig.\ 4f), although at low $\varepsilon$ a less reduction of ${\delta}\vec{r}^2(t)$ occurs, and one observes a relatively larger domain of anticyclonic motion (Fig.\ 4a-4c). 
We determine the diffusion coefficient $D$ of vortex motion, using the Green-Kubo formula \cite{Gr54,Ku57,ME07,HM13}: $D{=}\int_0^{\infty}{\langle}\vec{u}(0){\cdot}\vec{u}(t){\rangle}dt$, with $\vec{u}(t)$ being the velocity of vortex motion. Results for the forced cell are compared with the reference cell in the insets. We see that with large $\varepsilon$ ($\varepsilon{\ge}5.0$) data for the two cells overlap. With decreasing $\varepsilon$, $D$ decreases more rapidly in the forced cell. When $\varepsilon{\approx}1.0$, $D$ of anticyclonic (cyclonic) motion in the forced cell becomes one (two) order in magnitudes less than that in the reference cell. Further decrease in $\varepsilon$ leads to a stationary vortex distribution in the forced cell as shown in Fig.\ 1c.         

The flow pattern in rapidly rotating convection is characterized by columnar vortices that undergo stochastic horizontal motion. We have shown that when a periodically topographic structure is introduced on the heated surface, the vortex motion can be strictly controlled to form stationary convection patterns with prescribed symmetries.  We demonstrate that the new patterns arise through a dynamical process of imperfect bifurcation with the effects of external forcing well described by a Ginzburg-Landau-like model. 
In rapidly rotating convection it is predicted that  heat and mass are transported mainly by the columnar vortices \cite{Ve59, JLMW99, KA12}.  
Our experimental findings of a wide parameter regime to manipulate these vortices through topological forcing may thus enable new experimental approaches to examine and exploit local heat and fluid transports in rotating flows.
The richness of patterns and vortex dynamics in modulated rotating convection observed in this study may stimulate further theoretical and numerical investigations, and will contribute to understanding of pattern formation in non-equilibrium systems constrained by nonuniform boundary conditions. 
 

This work is supported by the Fundamental Research Funds for the Central Universities, and the National Science Foundation of China under Grant Nos. 92152105 and 11772235.


\end{document}